\documentclass[pre,twocolumn,showpacs,superscriptaddress]{revtex4}
\usepackage{graphicx}
\usepackage{amsmath,amssymb}
\usepackage{epstopdf}

\newcommand{\myav}[1]{\langle #1\rangle}

\newcommand{\mynav}[1]{\operatorname{E}({#1})}
\newcommand{\mynavprime}[1]{\operatorname{E}'({#1})}

\newcommand{\steady}{{\mathrm{ss}}}

\newcommand{\myvec}[1]{{\bf #1}}
\newcommand{\Pvec}{\myvec{P}}
\newcommand{\Amat}{\myvec{A}}
\newcommand{\Pveceq}{\Pvec^{\mathrm{eq}}}
\newcommand{\Ivec}{\myvec{1}}

\newcommand{\latin}[1]{{\itshape #1}}

\newcommand{\eg}{\latin{e.\,g.}}

\newcommand{\ie}{\latin{i.\,e.}}

\newcommand{\etal}{\latin{et al.}}

\newcommand{\Naive}{Na\"\i ve}

\newcommand{\codename}[1]{{\sc #1}}
\newcommand{\COPASI}{\codename{copasi}}
\newcommand{\MATLAB}{\codename{matlab}}
\newcommand{\ARPACK}{\codename{arpack}}
\newcommand{\OCTAVE}{\codename{octave}}

\begin{document}

\title{Steady-state parameter sensitivity in stochastic modeling via
  trajectory reweighting}

\author{Patrick B. Warren}

\affiliation{Unilever R\&D Port Sunlight, Quarry Road East, Bebington,
  Wirral, CH63 3JW, UK.}

\author{Rosalind J. Allen}

\affiliation{SUPA, School of Physics and Astronomy, The University of
  Edinburgh, The Kings Buildings, Mayfield Road, Edinburgh, EH9 3JZ,
  UK.}

\date{February 8, 2012}

\begin{abstract}
Parameter sensitivity analysis is a powerful tool in the building and
analysis of biochemical network models. For stochastic simulations,
parameter sensitivity analysis can be computationally expensive,
requiring multiple simulations for perturbed values of the
parameters. Here, we use trajectory reweighting to derive a method for
computing sensitivity coefficients in stochastic simulations without
explicitly perturbing the parameter values, avoiding the need for
repeated simulations.  The method allows the simultaneous computation of multiple sensitivity coefficients. Our approach recovers results originally
obtained by application of the Girsanov measure transform in the
general theory of stochastic processes [A. Plyasunov and A.~P. Arkin,
  J. Comp. Phys. {\bf221}, 724 (2007)].  We build on these results to
show how the method can be used to compute steady-state 
sensitivity coefficients from a single simulation run, and
we present various efficiency improvements. For models of biochemical
signaling networks the method has a particularly simple
implementation. We demonstrate its application to a signaling network
showing stochastic focussing and to a bistable genetic switch, and
present exact results for models with linear propensity functions.
\end{abstract}



\maketitle

\section{Introduction}
Parameter sensitivity analysis is one of the most important tools
available for modelling biochemical networks. Such analysis is particularly
crucial in systems biology, where models may have hundreds
of parameters whose values are uncertain.  Sensitivity analysis allows
one to rank parameters in order of their influence on network
behaviour, and hence to target experimental measurements towards
biologically relevant parameters and to identify possible drug
targets. For deterministic models, the adjunct ODE method provides an
efficient way to compute the local sensitivity of a model to small
changes in parameters.  For stochastic models, however, parameter
sensitivity analysis can be computationally intensive, requiring
repeated simulations for perturbed values of the parameters. Here, we
demonstrate a method, based on trajectory reweighting, for computing
local parameter sensitivity coefficients in stochastic kinetic
Monte-Carlo simulations without the need for repeated simulations.

Sensitivity analysis of biochemical network models may take a number
of forms. One may wish to determine how a model's behaviour changes as
a parameter is varied systematically within some range (a parameter
sweep), its dependence on the initial conditions of a simulation, or
its sensitivity to changes in the structure of the model itself
(alternate mode-of-action hypotheses).  In this paper, we focus on the
computation of local parameter sensitivity coefficients. These
coefficients describe how a particular output $f$ of the model varies
when the $\alpha$-th parameter of the model, $k_\alpha$, is varied by
an infinitesimal amount, $k_\alpha \to k_\alpha +h$:
\begin{equation}
\frac{\partial f}{\partial k_\alpha} 
= \lim_{h\to0} \frac{f'-f}{h}\,.
\end{equation}
where $f'$ is the output of the model computed in a system with
$k_\alpha$ changed to $k_\alpha +h$. For deterministic models, where
the dynamics of the variables $x_i(t)$ can be described by a set of
deterministic ordinary differential equations (ODEs) ${\partial
  x_i}/{\partial t}=g(x_i, k_\alpha)$, differentiation of the ODEs
with respect to $k_\alpha$ shows that the sensitivity coefficients
$C_{i,\alpha}\equiv\partial x_i/\partial k_\alpha$ obey an
\emph{adjunct} set of ODEs,
\begin{equation}
\frac{\partial C_{i,\alpha}}{\partial t}
=\sum_j \frac{\partial g}{\partial x_j} C_{j,\alpha}
+\frac{\partial g}{\partial k_\alpha}\,.
\label{eq:1}
\end{equation}
These adjunct ODEs can be integrated alongside the original ODEs to
compute the sensitivity coefficients ``on the fly'' in a deterministic
simulation of a biochemical network.

Stochastic models of biochemical networks are (generally)
continuous-time Markov processes \cite{vK92} which are solved
numerically by kinetic Monte-Carlo simulation, using standard methods
such as the Gillespie \cite{Gil77} or Gibson-Bruck \cite{GB00}
algorithms. Replicate simulations will produce different trajectories;
we wish to compute how the {\em{average}} value $\myav{f}$ of some function
$f(t)$ of the model changes with the parameter $k_\alpha$:
\begin{equation}
\frac{\partial \myav{f}}{\partial k_\alpha} 
= \lim_{h\to0} \frac{\myav{f}'-\myav{f}}{h}\,.
\label{eq:hlim}
\end{equation}
where the averages are taken across replicate simulation runs. If one
is interested in steady-state (\ie\ time-independent) parameter
sensitivities, the averages in Eq.~\eqref{eq:hlim} may instead be time
averages taken over a single simulation run. \Naive\ evaluation of
parameter sensitivities via Eq.~\eqref{eq:hlim} is very inefficient,
since one is likely to be looking for a small difference between two
fluctuating quantities.  There are several existing approaches that
get around this problem: spectral methods \cite{KDN07}, a method based
on the Girsanov measure transform \cite{PA07}, and methods which
re-use the random number streams \cite{RSK10}. In this paper, we
develop a method based on trajectory reweighting, which is simple to
implement in existing kinetic Monte Carlo codes and provides a way to
compute steady-state parameter sensitivity coefficients ``on-the-fly''
in stochastic simulations of biochemical networks.  The method
provides an accessible alternative to the Girsanov measure transform
pioneered by Plyasunov and Arkin \cite{PA07}.  Indeed several of our
equations in Section \ref{sec:tw} are equivalent to those
Ref.~\onlinecite{PA07}.  However, we go beyond previous work
by showing in practical terms how the method can be implemented in
standard stochastic simulation algorithms, extending the method to the
computation of parameter sensitivities in the steady state, and
showing how time-step preaveraging can be used to improve the
efficiency of the calculations.

\section{Trajectory reweighting}
\label{sec:tw}
The basic idea behind trajectory reweighting is as follows. In a
kinetic Monte-Carlo simulation, for a given set of parameters any
given trajectory has a statistical weight which measures the
probability that it will be generated by the algorithm \cite{HS07,
  KM08, RGP10}; this weight can be expressed as an analytical function
of the states of the system along the trajectory and of the parameter
set. This analytical function also allows us to compute the  statistical weight
 for this {\em{same}} trajectory, in a system with a  {\em{different}} set of
parameters: \ie\  its weight in the ensemble of trajectories
with perturbed parameters.  This allows us
in principle to compute the average $\myav{f}'$ in Eq.~\eqref{eq:hlim}
for the perturbed parameter set, using only a set of trajectories
generated with the unperturbed parameter set. For most applications
this is inefficient, because the weight of a trajectory in the
perturbed ensemble is typically very low, resulting in poor
sampling. However, it turns out that trajectory reweighting does
provide an effective way to compute local parameter sensitivity
coefficients.

\subsection*{Trajectory reweighting for kinetic Monte-Carlo simulations}
More specifically, let us consider a typical implementation of the
Gillespie algorithm \cite{Gil77} (similar arguments apply to more
recent algorithms, such as Gibson-Bruck \cite{GB00}). Here, the state
of the system is characterised by a set of discrete quantities $N_i$,
typically representing the number of molecules of chemical species
$i$.  Transitions between states are governed by propensity functions
$a_\mu(N_i, k_\alpha)$ where $\mu$ labels the possible reaction
channels and the quantities $k_\alpha$ are parameters in the problem,
typically reaction rates ($k_\alpha$ represents the $\alpha$-th such parameter). A kinetic Monte-Carlo trajectory is
generated by stepping through the space of states $N_i$ in the
following way.  We first compute the propensity functions $a_\mu(N_i)$
for all the possible transitions out of the current state.  We then
choose a time step (\ie\ waiting time) $\delta t$ from an exponential
distribution $p(\delta t)=\tau^{-1}e^{-\delta t/\tau}$, where the
state-dependent mean timestep (the mean waiting time before exiting
the current state) is $\myav{\delta t}=\tau\equiv(\sum_\mu
a_\mu)^{-1}$.  We choose a reaction channel with probability
$p_\mu=a_\mu/\sum_\mu a_\mu$.  We advance time by $\delta t$ and
update the values of $N_i$ according to the chosen reaction channel.
We are now in a new state, and the above steps are repeated.

Now let us consider the statistical weight of a given trajectory
generated by this algorithm \cite{HS07, KM08, RGP10}. In each step,
the probability of choosing the value of $\delta t$ that we actually
chose is proportional to $\tau^{-1}e^{-\delta t/\tau}$, and the
probability of choosing the reaction channel that we actually chose is
equal to $a_\mu/\sum_\mu a_\mu$. We can therefore associate a weight
$P$ with the whole trajectory, which is proportional to the
probability of generating the sequence of steps which we actually
generated:
\begin{equation}
\begin{array}{ll}
P&=\textstyle{\prod_{\mathrm{steps}}}\,
(a_\mu/\sum_\mu a_\mu)\times (\tau^{-1}e^{-\delta t/\tau})\\[6pt]
&=\textstyle{\prod_{\mathrm{steps}}}\,
a_\mu \,e^{- (\sum a_\mu) \delta t}\,.
\end{array}
\label{eq:Pdef}
\end{equation}
The second line follows by eliminating $(1/\sum_\mu a_\mu)\times
\tau^{-1}\equiv 1$ (note that because Eq.~\eqref{eq:Pdef} is not
normalized, $P$ is a \emph{weight} rather than a true probability).

In a typical kinetic Monte-Carlo simulation, we generate multiple independent
trajectories of length $t$, for a given parameter set. The probability
of generating any given trajectory in this sample will be proportional
to its weight $P$, defined in Eq.~\eqref{eq:Pdef}. We then compute
the average $\myav{f(t)}$ of some function $f(t)$ of the state of the
system by summing over the values of $f$, at time $t$, for these
trajectories.

Having generated this set of trajectories, let us now suppose we wish
to re-use them to compute the average $\myav{f(t)}'$ which we would
have obtained had we repeated our simulations for some {\em{other}}
parameter set. It turns out that we can compute this average by
summing over the same set of trajectories, multiplied by the ratio of
their statistical weights for the perturbed and unperturbed parameter
sets. To see this, we first recall that an average,
\eg\ $\myav{f(t)}$, can be written as a sum over all {\em{possible}}
trajectories $j$ of length $t$, multiplied by their statistical
weights $P_j$: $\myav{f(t)} = (\sum_j P_j f_j(t))/(\sum_j
P_j)$. Writing the perturbed average $\myav{f(t)}'$ in this way, we
obtain
\begin{equation}
\textstyle
\myav{f(t)}'= \frac{\sum_j P_j' f_j(t)}{\sum_j P_j'}
= \frac{\sum_j (P_j'/P_j) P_j f_j(t)}{\sum_j
  (P_j'/P_j)P_j}
= \frac{\myav{f(t) {P'}\!/{P}}}  {\myav{{P'}\!/{P}}}
\label{eq:pppp}
\end{equation}
%
%
where $P$ and $P'$ are the trajectory weights (calculated using
Eq.~\eqref{eq:Pdef}) for the original and perturbed models
respectively. In another context, Eq.(\ref{eq:pppp}) has been used to reweight trajectory statistics in order to sample rare events in biochemical networks \cite{RGP10}; it also forms the basis of umbrella sampling methods for particle-based Monte Carlo simulations \cite{frenkelbook}.

Whilst in principle  Eq.(\ref{eq:pppp}) provides a completely general way to transform between trajectory ensembles with different parameter sets, in practice it
is useless for any significant deviation of the parameter set from the
original values, for two reasons.  First, the statistical errors in
the computation of $\myav{f}'$ grow catastrophically with the size of
the perturbation, because the original trajectories become
increasingly unrepresentative of the perturbed model.  Second, the
computational cost of determining the trajectory weights for the
perturbed and unperturbed parameter sets via Eq.~\eqref{eq:Pdef} is
only marginally less than the cost of computing $\myav{f}'$ directly
by generating a new set of trajectories for the perturbed parameter
set.

\subsection*{Computation of parameter sensitivity coefficients}
It turns out, however, that Eq.~\eqref{eq:pppp} is useful for the
computation of parameter sensitivity coefficients, where the deviation
between the original and perturbed parameter sets is
infinitesimal. Let us suppose that the perturbed problem corresponds
to a small change in a single parameter, such as $k_\alpha'=
k_\alpha+h$; the corresponding sensitivity coefficient is defined by
Eq.~\eqref{eq:hlim}. As we show in Supplementary Material Section
\ref{sec:rat}, differentiating Eq.~\eqref{eq:pppp}
with respect to $k_\alpha'$ leads to the following expression for the
sensitivity coefficient:
\begin{equation}
\frac{\partial\myav{f}}{\partial k_\alpha}=
\myav{f W_{k_\alpha}} - \myav{f}\myav{W_{k_\alpha}}\,.
\label{eq:key1}
\end{equation}
where 
\begin{equation}
W_{k_\alpha}=\frac{1}{P}\frac{\partial P}{\partial k_\alpha}=%
\frac{\partial\ln P}{\partial k_\alpha}\,
\label{eq:key2}
\end{equation}
Supplementary Material Section \ref{sec:rat} also shows how to
generalize this approach to higher-order derivatives.
Combining Eq.~\eqref{eq:Pdef} with Eq.~\eqref{eq:key2} shows that the
``weight function'' $W_{k_\alpha}$ can be expressed as a sum over all
steps in the trajectory:
\begin{equation}
W_{k_\alpha}
={\textstyle \sum_{\mathrm{steps}}}\,
\delta W_{k_\alpha}
\label{eq:key2aa}
\end{equation}
where
\begin{equation}
\delta W_{k_\alpha}=\frac{\partial\ln a_{\mu}}{\partial k_\alpha}
 - \frac{\partial(\sum a_{\mu})}{\partial k_\alpha} \delta t\,.
\label{eq:key2a}
\end{equation}

Eqs.~\eqref{eq:key1}--\eqref{eq:key2a} are the key results of this
paper, since they point to a practical way to compute parameter
sensitivity coefficients in kinetic Monte-Carlo simulations. To
evaluate the (time-dependent) parameter sensitivity $\partial
\myav{f(t)} / \partial k_\alpha$, one tracks a weight function
$W_{k_\alpha}(t)$, which evolves according to Eqs.~\eqref{eq:key2aa}
and \eqref{eq:key2a}. One also tracks the function $f(t)$ of interest.
The covariance between $W_{k_\alpha}$ and $f$, at the time $t$ of
interest, computed over multiple simulations, then gives the
sensitivity of $\myav{f(t)}$ to the parameter in question (as in
Eq.~\eqref{eq:key1}). Tracking $W_{k_\alpha}$ should be a
straightforward addition to standard kinetic Monte-Carlo
schemes. Moreover we note that $f$ could be any function of the
variables of the system---for example, if one were interested in the
parameter sensitivity of the noise in particle number $N_i$, one could
choose $f(N_i)=N_i^2$. More complex functions of the particle numbers,
involving multiple chemical species, could also be used (see examples
below).

This prescription for computing parameter sensitivities presents,
however, some difficulties in terms of statistical sampling. The two
terms in Eq.~\eqref{eq:key2a} are statistically independent quantities
with the same expectation value, $\partial\ln(\sum a_{\mu})/\partial
k_\alpha$.  Hence they cancel on average but the variances add.  Thus
we expect that $W_{k_\alpha}$ is a stochastic process with a zero
mean, $\myav{W_{k_\alpha}}=0$, and a variance that should grow
approximately linearly with time---as shown for a simple example case
in Supplementary Material Section \ref{sec:linear}---in effect
$W_{k_\alpha}$ behaves as a random walk (\ie\ a Wiener process).  In
terms of controlling the sampling error, this means that the number of
trajectories over which the covariance is evaluated should increase in
proportion to the trajectory length, since the standard error in the
mean is expected to go as the square root of the variance divided by
the number of trajectories \cite{note-error}. In Section \ref{sec:ss},
we discuss a way to avoid this problem, when computing steady-state
parameter sensitivities.

\subsection*{Simplifications and practical implementation}
Without loss of generality we can presume that the parameter $k_\alpha$
will appear in only one of the propensity functions, which we call
$a_\alpha$ \cite{note-epar}.  With this presumption,
Eq.~\eqref{eq:key2a} becomes
\begin{equation}
\delta W_{k_\alpha}=\frac{\partial\ln a_{\alpha}}{\partial k_\alpha}
(\delta_{\mu\alpha} - a_{\alpha} \delta t).
\label{eq:key2b}
\end{equation}
Eq.~\eqref{eq:key2b} makes a direct link with the Girsanov measure
transform method introduced by Plyasunov and Arkin, being essentially
the same as Eq.~(31b) in Ref.~\onlinecite{PA07}.

A further simplification occurs if $k_\alpha$ is the rate coefficient
of the $\alpha$-th reaction, so that $a_\alpha$ is linearly
proportional to $k_\alpha$.  One then has $\partial \ln
a_\alpha/\partial k_\alpha=1/k_\alpha$ and Eq.~\eqref{eq:key2aa}
becomes
\begin{equation}
W_{k_\alpha}= \frac{1}{k_\alpha}\left[Q_\alpha-{\textstyle\sum_{\mathrm{steps}}}\,
a_{\alpha}\delta t \right]
\label{eq:key2c}
\end{equation}
where $Q_\alpha$ counts the number of times that the $\alpha$-th
reaction is visited.  This is essentially the same as Eq.~(9b) of
Plyasunov and Arkin's work, Ref.~\onlinecite{PA07}.

Eq.~\eqref{eq:key2c} suggests a very simple way to implement parameter
sensitivity computations in existing kinetic Monte-Carlo codes. One
simply modifies the chemical reaction scheme such that each reaction
whose rate constant is of interest generates a ``ghost'' particle in
addition to its other reaction products (this is similar to the clock
trick in Ref.~\onlinecite{WHD+06}).  There should be a different
flavour of ghost particle for each reaction of interest, and ghost
particles should not participate in any other reactions.  $Q_\alpha
(t)$ is then simply given by the number of ghost particles associated
with the $\alpha$-th reaction which are present at time $t$.  In
Section \ref{sec:egs}, we use this approach to compute sensitivity
coefficients using the \emph{unmodified} \COPASI\ \cite{HS06}
simulation package.  In Supplementary Material Section
\ref{sec:linear} we also exploit this trick to obtain some exact
results for linear propensity functions.

\section{Steady state}
\label{sec:ss}
So far, we have discussed the computation of time-dependent parameter
sensitivity coefficients $\partial \myav{f(t)}/\partial k_\alpha$, by
evaluating the covariance of the weight function $W_{k_\alpha}(t)$
with the function $f(t)$ over multiple simulation runs. Often,
however, one is interested in the parameter sensitivity of the
{\em{steady-state}} properties of the system $\partial
\myav{f}_\steady/\partial k_\alpha$; this is a time-independent
quantity. We now discuss the computation of steady-state parameter
sensitivities using trajectory reweighting. We show that in this
case, first, the problem of poor sampling of $W_{k_\alpha}(t)$ for
long times can be circumvented, second, one can obtain sensitivity
coefficients from a single simulation run, and third, one can
improve efficiency by a procedure which we call time-step
pre-averaging.

\subsection*{The ensemble-averaged correlation function method}
To compute steady-state parameter sensitivities, one might imagine
that we could simply apply the method discussed in Section
\ref{sec:tw}, taking the limit of long times, when the system should
have relaxed to its steady state. However, this does not work, because
the variance between trajectories of the weight function
$W_{k_\alpha}$ increases in time, making it impossible to obtain good
statistics at long times.  To circumvent this problem, we note that
the right hand side of Eq.~\eqref{eq:key1} is unaltered if
$W_{k_\alpha}$ is offset by a constant. Thus we may write the
parameter sensitivity in the form of a two-point time-correlation
function:
\begin{equation}
\begin{array}{l}
\displaystyle
\frac{\partial \myav{f(t)}}{\partial k_\alpha} = C(t,t_0) = %
\myav{f(N_i,t)\, \Delta W_{k_\alpha}(t,t_0)}\\[3pt]
{}\hspace{8em}- \myav{f(N_i,t)}\,\myav{\Delta W_{k_\alpha}(t,t_0)}
\end{array}
\label{eq:cdef}
\end{equation}
where 
\begin{equation}
\Delta W_{k_\alpha}(t,t_0)=W_{k_\alpha}(t)- W_{k_\alpha}(t_0),
\label{eq:DWt}
\end{equation}
and $t_0$ is some arbitrary reference time such that $t-t_0=\Delta
t>0$. This relation has the advantage that we may choose $\Delta t$
sufficiently small to make the variance of $\Delta
W_{k_\alpha}(t,t_0)$ manageable. Importantly, in steady-state conditions,
we expect that the correlation function depends only on the time
difference and not separately on the two times, so that
$C(t,t_0)=C(\Delta t)$, with $C(0)=0$ and $C(\Delta
t)\to\partial\myav{f}_\steady/\partial k_\alpha$ as $\Delta t\to\infty$.
Thus to calculate the sensitivity coefficient under steady state
conditions, all we need to do is compute the steady-state correlation
function defined in Eq.~\eqref{eq:cdef}, choosing a suitable
``reference'' time $t_0$ when the system is already in the
steady-state, then take the asymptotic (large $\Delta t$) value of
this correlation function.  We expect the correlation function to
approach its asymptotic value on a timescale governed by the (likely
short) relaxation time spectrum in the steady state, so that for most
problems large values of $\Delta t$ should not be required. Noting
that in this method, as in Section \ref{sec:tw}, the averages in
Eq.~\eqref{eq:cdef} are computed over multiple independent simulation
runs, we term this approach the \emph{ensemble-averaged correlation
  function method}.

From a practical point of view, this method involves the following set
of steps or `recipe':
\begin{enumerate}
\item{Choose two time points $t_1$ and $t_2$ such that the system has
  already reached its steady state at time $t_1$ and $t_2 = t_1 +
  \Delta t$ where $\Delta t$ is greater than the typical relaxation
  time of the quantity $f$ of interest (typically this is the same as
  the longest relaxation time in the system as a whole).}
\item{Compute $W_{k_\alpha}$ at times $t_1$ and $t_2$ and $f$ at time
  $t_2$.}
\item{Calculate the difference $\Delta W_{k_\alpha} =
  W_{k_\alpha}(t_2)-W_{k_\alpha}(t_1)$. Compute also the product
  $f(t_2)\Delta W_{k_\alpha}$.}
\item{Repeat steps 1-3 for many independent simulation runs and
  compute the averages $\myav{f(t_2)}$, $\myav{\Delta
    W_{k_\alpha}(t_2-t_1)}$ and $\myav{f(t_2) \Delta
    W_{k_\alpha}(t_2-t_1)}$ over the replicate simulations.}
\item{Calculate the correlation function $C(\Delta t)=\myav{f \Delta
    W_{k_\alpha}}-\myav{ f}\myav{\Delta W_{k_\alpha}}$. As long as
  $\Delta t$ is large enough this provides a measurement of
  $\partial\myav{f}_\steady/\partial k_\alpha $.}
\end{enumerate}

\subsection*{The time-averaged correlation function method}
It turns out, however, that for steady-state parameter sensitivities,
we do not need to average over multiple simulation runs---we can
instead compute time-averages over a single simulation run. This
amounts to replacing the steady-state ensemble averaged sensitivity
$\partial \myav{f}_\steady/\partial k_\alpha$ by the \emph{time averaged}
version $\partial \overline{f}_\steady/\partial k_\alpha$, where
\begin{equation}
\overline{f}=\frac{\textstyle\sum_{\mathrm{steps}} \,f\delta t}%
{\sum_{\mathrm{steps}} \delta t}
\label{eq:tave}
\end{equation}
(recalling that in kinetic Monte Carlo, the timestep $\delta t$ varies
between steps). In principle, $\partial \overline{f}_\steady/\partial
k_\alpha$ could be obtained by computing the time-averaged version of
Eq.~\eqref{eq:cdef}:
\begin{equation}
C_{\mathrm{time\,\,av}}(\Delta t) = \overline{f(t)\, %
\Delta W_{k_\alpha}(t,t_0)}- \overline{f(t)}\,\,%
\overline{\Delta W_{k_\alpha}(t,t_0)}
\label{eq:cdefta}
\end{equation}
and taking the limit of large $\Delta t = t-t_0$. Eq.~\eqref{eq:cdefta}
requires one to keep track of $W_{k_\alpha}$ a precise time $\Delta t$
in the past; since the time step is not constant in kinetic Monte
Carlo, this is rather inconvenient to implement.

Fortunately, however, tracking the weight function at a precise time
in the past turns out to be unnecessary. As $\Delta t$ becomes large,
the stochastic differences between individual time steps cancel out and
it becomes equivalent simply to compute the average
\begin{equation}
C_{\mathrm{time\,\,av}}(n) = \overline{f(t)\, \Delta W_{k_\alpha}(n)}%
- \overline{f(t)}\,\,\overline{\Delta W_{k_\alpha}(n)}
\label{eq:cdeftan}
\end{equation}
where
\begin{equation}
\Delta W_{k_\alpha}(n)=W_{k_\alpha}(t)-W_{k_\alpha}(\hbox{$n$
  steps ago}).
\label{eq:DWn}
\end{equation}
and to use the fact that
$C_{\mathrm{time\,\,av}}(n)\to\partial\overline{f}_\steady/\partial
k_\alpha$ as $n\to\infty$. One can quite easily keep track of $\Delta
W_{k_\alpha}(n)$, for instance by maintaining a circular history array
storing $W_{k_\alpha}$ over the last $n$ steps. This approach, which
we denote the {\em{time-averaged correlation function method}}, has
the important advantage that one can obtain the steady state parameter
sensitivity from a single simulation run. 

The recipe for using the time-averaged correlation function method is then:
\begin{enumerate}
\item{Choose a time interval $\Delta t$ which is greater than the
  typical relaxation time of the quantity $f$ of interest. Estimate
  the typical number of steps $n$ taken in time $\Delta t$: $n =
  \Delta t/\bar{\tau}$.}
\item{For a simulation of the system in the steady state, record $f$
  and $W_{k_\alpha}$ every $n$ steps (we denote each of these
  recordings a `timeslice').}
\item{For each timeslice $i$, compute the difference between
  $W_{k_\alpha}^{(i)}$ and its value in the previous timeslice:
  $\Delta W_{k_\alpha}^{(i)} =
  W_{k_\alpha}^{(i)}-W_{k_\alpha}^{(i-1)}$. Compute also
  $f^{(i)}\Delta W_{k_\alpha}^{(i)}$.}
\item{Compute the averages over all timeslices of $f$, $\Delta
  W_{k_\alpha}$ and $f\Delta W_{k_\alpha}$.}
\item{Calculate the correlation function $C(n) = \overline{f \Delta
    W_{k_\alpha}}-(\overline{ f})(\overline{\Delta W_{k_\alpha}})$. As
  long as $n$ is large enough this provides a measurement of
  $\partial\overline{f}_\steady/\partial k_\alpha $.}
\end{enumerate}

\subsection*{Time-step pre-averaging}
The time-averaged correlation function method is a convenient way to
compute parameter sensitivities in a standard kinetic Monte Carlo
scheme, in which both a new timestep and a new reaction channel are
chosen stochastically at every step. However, choosing a new time step
at every iteration is computationally expensive since it requires a
random number, and is not strictly necessary for the computation of
steady-state parameter sensitivity coefficients. Improved efficiency
can be achieved by choosing only the new reaction channel
stochastically at each iteration, and replacing $\delta t$ by the mean
timestep $\tau\equiv(\sum_\mu a_\mu)^{-1}$ corresponding to the
current state (note that this is state dependent since it depends on
the propensity functions). This amounts to \emph{pre-averaging} over
the distribution of possible time steps for a given state of the
system.  It can be proved formally that if we run our simulations for
a sufficiently long time, Eq.~\eqref{eq:tave} is equivalent to
\begin{equation}
\overline {f} = 
\frac{\sum_{\mathrm{steps}}\,f\tau}{
\sum_{\mathrm{steps}}\,\tau}\,.
\label{eq:pre}
\end{equation}  
Intuitively, this relation arises because a
sufficiently long trajectory, under steady state conditions, will
visit each state an arbitrarily large number of times and thus
thoroughly sample the distribution of waiting times in each state.

One cannot, however, compute the parameter sensitivity
$\partial\overline{f}_\steady/\partial k_\alpha$ simply by evaluating the
time averages in Eq.~\eqref{eq:cdefta} or \eqref{eq:cdeftan} using the
new definition, Eq.~\eqref{eq:pre}. This is because $\tau$ itself
depends on the parameter $k_\alpha$. Instead, a slightly more
complicated expression for $\partial\overline{f}_\steady/\partial
k_\alpha$ is required; this is given in Supplementary Material Section
\ref{sec:preav}. Thus, time-step pre-averaging
provides a more efficient way to compute the steady-state parameter
sensitivity (since the time does not need to be updated in the Monte
Carlo algorithm), at the cost of a slight increase in mathematical
complexity.

\section{Examples}
\label{sec:egs}
We now apply the methods described above to three case studies: a
model for constitutive gene expression for which we can compare our
results to analytical theory, a simple model for a signaling pathway
with stochastic focussing, and a model for a bistable genetic switch.
The second and third examples are chosen because they exhibit the kind
of non-trivial behaviour found in real biochemical networks, yet the
state space is sufficiently compact that the parameter sensitivities
can be checked using finite-state projection (FSP) methods
\cite{MK06}.  Our implementation of the FSP methods is described more
fully in Supplementary Material Section \ref{sec:fsp}.

\begin{figure}
\begin{center}
\includegraphics{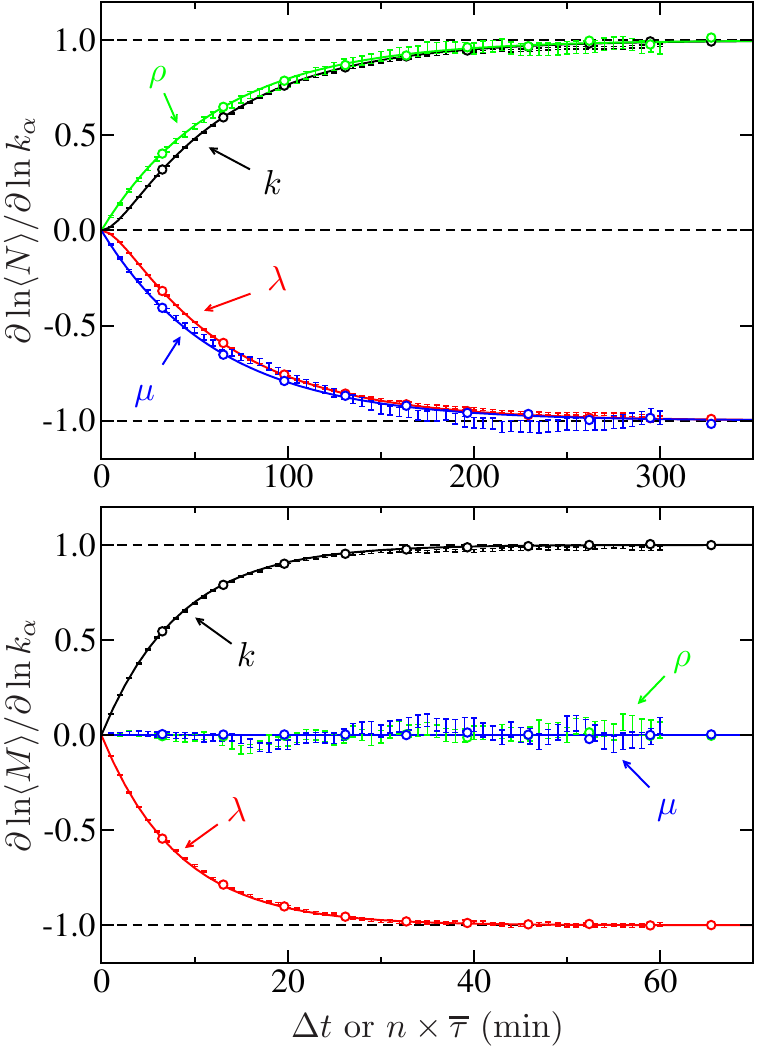}
\end{center}
\caption[?]{(color online) Time correlation functions $C(\Delta t)$
  for the sensitivity of the average protein number $\myav{N}_\steady$ (top)
  and the average mRNA number $\myav{M}_\steady$ to the model parameters, for
  the constitutive gene expression model in Section \ref{sec:santmac}.
  Points with error bars are simulations using the ensemble-average
  correlation function method; error bars are estimated by
  block-averaging (100 blocks of $10^3$ trajectories; total
  $\approx10^{10}$ time steps).  Open circles are simulations using
  the time-average correlation function method with time step
  pre-averaging (plotted as a function of $n\bar{\tau}$); results are
  averages over 10 trajectories each of length $10^9$ steps (in this
  case the error bars are smaller than symbols).  Solid lines are
  theoretical predictions from Eq.~\eqref{eq:santmaccf}.  Parameters
  are $k=2.76\,\mathrm{min}^{-1}$, $\lambda=0.12\,\mathrm{min}^{-1}$,
  $\rho=3.2\,\mathrm{min}^{-1}$, $\mu=0.016\,\mathrm{min}^{-1}$,
  corresponding to the {\em{cro}} gene in a recent model of phage
  lambda \cite{SM04}. For these parameters $\myav{M}_\steady=23$ and
  $\myav{N}_\steady=4600$.\label{fig:corrn}}
\end{figure}

\subsection{Constitutive gene expression}
\label{sec:santmac}
We first consider a simple stochastic model for the expression of a
constitutive (unregulated) gene, represented by the following chemical
reactions:
\begin{equation}
\emptyset\overset{k}\to \mathrm{M} \overset{\lambda}\to\emptyset,\qquad
\mathrm{M}\overset{\rho}\to \mathrm{M}+\mathrm{N},\qquad
\mathrm{N}\overset{\mu}\to\emptyset.
\end{equation}
Here, M represents messenger RNA (synthesis rate $k$, degradation rate
$\gamma$) and N represents protein (synthesis rate $\rho$, degradation
rate $\mu$).  This model has linear propensities (as defined in
Supplementary Material Section \ref{sec:linear}),
which implies that the mean copy numbers $\myav{M(t)}$ and
$\myav{N(t)}$ of mRNA and protein respectively obey the chemical rate
equations
\begin{equation}
\frac{d\myav{M}}{dt}=k-\lambda\myav{M},\quad
\frac{d\myav{N}}{dt}=\rho\myav{M}-\mu\myav{N}
\label{eq:santmac}
\end{equation}
from which follow the steady state mean copy numbers: 
\begin{equation}
\myav{M}_\steady=\frac{k}{\lambda}\,,\quad
\myav{N}_\steady=\frac{\rho k}{\lambda\mu}\,.
\label{eq:santmacss}
\end{equation}
For this problem, steady-state sensitivity coefficients can be
computed analytically by taking derivatives of
Eqs.~\eqref{eq:santmacss} with respect to the parameters of
interest. Moreover, as shown in Supplementary Material Section
\ref{sec:linear}, explicit expressions can also be found for the
components of the correlation functions defined by
Eqs.~\eqref{eq:cdef} and \eqref{eq:DWt}:
\begin{equation}
\begin{array}{l}
\myav{M \Delta W_{\ln k}}_\steady=-\myav{M \Delta W_{\ln\lambda}}_\steady
=\myav{M}_\steady(1-e^{-\lambda\Delta t}),\\[6pt]
\myav{M \Delta W_{\ln\rho}}_\steady=\myav{M \Delta W_{\ln\mu}}_\steady=0,\\[6pt]
\myav{N \Delta W_{\ln k}}_\steady=-\myav{N \Delta W_{\ln\lambda}}_\steady\\
{}\hspace{1em}=\myav{N}_\steady
[{\lambda(1-e^{-\mu\Delta t})-\mu(1-e^{-\lambda\Delta t})}]/
({\lambda-\mu}),\\[6pt]
\myav{N \Delta W_{\ln\rho}}_\steady=-\myav{N \Delta W_{\ln\mu}}_\steady
=\myav{N}_\steady(1-e^{-\mu\Delta t}),
\end{array}
\label{eq:santmaccf}
\end{equation}
where for notational convenience we consider the sensitivity with
respect to the logarithm of the parameter value (\eg\ $W_{\ln k} 
\equiv kW_k$).

Figure \ref{fig:corrn} shows the time correlation functions of
Eqs.~\eqref{eq:cdef} and \eqref{eq:DWt}, computed over multiple
stochastic simulation runs using the ensemble-averaged correlation
function method, together with the analytical results of
Eq.~\eqref{eq:santmaccf} (solid lines). The agreement between the
analytic theory and simulation results is excellent. The time
correlation functions converge to the expected steady state
sensitivity coefficients (horizontal lines in
Fig.~\ref{fig:corrn}). For the protein correlation functions
(${\partial \ln \myav{N}_\steady}/{\partial \ln k_\alpha}$), this occurs on a
timescale governed by the relaxation rate of protein number
fluctuations $1/\mu \approx 60$\,min, while the mRNA correlation
functions (${\partial \ln \myav{M}_\steady}/{\partial \ln k_\alpha}$) reach
their asymptotic values on a timescale governed by the mRNA decay rate
$1/\lambda \approx 8$\,min.

Figure \ref{fig:corrn} (open circles) also shows the same correlation
functions, computed instead from a single stochastic simulation run,
using the time-averaged correlation function method, with time-step
pre-averaging. Although this method gives correlation
functions (Eqs.~\eqref{eq:cdeftan} and \eqref{eq:DWn}) in terms of the
number of steps $n$ in the history array, rather than the time
difference $\Delta t$, these can be converted to time correlation
functions by multiplying $n$ by the expected \emph{global} mean time
step $\overline{\tau}$ (the average over states of the state-dependent
mean time step $\tau$).

Comparing the results of the ensemble-averaged and time-averaged
correlation function methods in Fig.~\ref{fig:corrn} we see that the
two methods give essentially the same results, but the time-averaged
method produces greater accuracy (smaller error bars), for the same
total number of simulation steps. Moreover, because we have used
time-step pre-averaging with the time-averaged correlation function
method, each simulation step is computed approximately twice as fast
as in the original kinetic Monte Carlo algorithm, since one does not
need to generate random numbers for the time steps
\cite{note-efficiency}.

\subsection{Stochastic focusing}
\label{sec:stochf}
We now turn to a more sophisticated case study, based on the
stochastic focusing model of Paulsson \etal\ \cite{PBE00}. In this
biochemical network, a input signal molecule S downregulates the
production of an output signal (or response) molecule R. Stochastic
fluctuations play a crucial role, making the output much more
sensitive to changes in the input than would be predicted by a
deterministic (mean-field) model.

Our reaction scheme, given in Eq.~\eqref{eq:sf}, contains just two
chemical species, S and R. The production and degradation of S (the
input signal) are straightforward Poisson processes with rates $k_s$
and $k_d$ respectively.  The production of R (the output signal) is
negatively regulated by S, and its degradation rate is set to unity to
fix the time scale.  Thus we have
\begin{equation}
\emptyset\overset{k_s}\to \mathrm{S} \overset{k_d}\to\emptyset\,,\qquad
\emptyset\overset{k^\star}\to \mathrm{R} \overset{1}\to\emptyset\,,
\label{eq:sf} 
\end{equation}
where we use a Michaelis-Menten-like form to represent the negative
regulation: $k^\star={k_p}/({S+K})$, with $S$ being the copy number of
the input signal molecule.  Taking a mean-field approach, we might
suppose that the average copy numbers $\myav{S}$ and $\myav{R}$ should
obey the chemical rate equations
\begin{equation}
\frac{d\myav{S}}{dt}=k_s-k_d\myav{S},\quad
\frac{d\myav{R}}{dt}=\frac{k_r}{\myav{S}+K}-\myav{R}\,.
\label{eq:stochf}
\end{equation}
and that therefore the steady state copy numbers should be given by
\begin{equation}
\myav{S}_\steady=\frac{k_s}{k_d}\,,\quad
\myav{R}_\steady=\frac{k_r}{\myav{S}_\steady+K}\,.
\label{eq:stochfmft}
\end{equation}
In reality, while Eq.~\eqref{eq:stochfmft} gives the correct result
for the mean input signal $\myav{S}_\steady$, it is manifestly
\emph{incorrect} for the mean output signal $\myav{R}_\steady$. For
example for
\begin{equation}
k_s=500,\quad k_d=100,\quad k_r=900,\quad K=0.09
\label{eq:stochfpar}
\end{equation}
we find from kinetic Monte Carlo simulations $\myav{S}_\steady=5$, as
predicted by Eq.~\eqref{eq:stochfmft}, but $\myav{R}_\steady\approx 290$,
whereas Eq.~\eqref{eq:stochfmft} predicts ${k_r}/({\myav{S}_\steady+K})=
176.82$. This failure of the mean-field prediction arises because of
the non-linearity of the Michaelis-Menten-like form of the production
propensity for R.

\begin{figure}
\begin{center}
\includegraphics{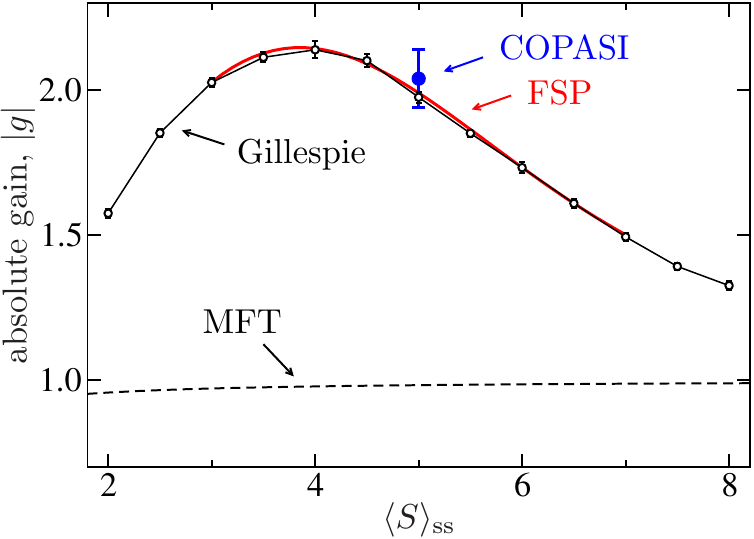}
\end{center}
\caption[?]{(color online) Steady state absolute differential gain for
  the stochastic focusing model in Section \ref{sec:stochf}, where
  $k_s$ is varied to control $\myav{S}_\steady$, with other parameters as in
  Eq.~\eqref{eq:stochfpar}.  Open circles are Gillespie simulations
  using the time-average correlation function method with time step
  pre-averaging; error bars are estimated by averaging over 10
  trajectories of length $10^8$ steps.  The history array length was
  $n=2\times10^4$.  The filled circle (blue) is from a Gibson-Bruck
  simulation in \COPASI\ using the ensemble-average correlation
  function method; error bars are from block averaging (10 blocks of
  $10^3$ samples).  The thick solid line (red) is the numerical result
  from the finite state projection (FSP) algorithm.  The dashed line
  is the mean-field theory (MFT) prediction.\label{fig:stochf}}
\end{figure}

Our aim is to compute the \emph{differential gain},
\begin{equation}
g=\frac{\partial\ln\myav{R}_\steady}{\partial\ln\myav{S}_\steady}\,.
\end{equation}
which describes the local steepness of the signal-response relation
($\myav{R}_\steady\sim\myav{S}_\steady^g$). The gain measures the sensitivity of the system's output $\myav{R}_\steady$ to its input $\myav{S}_\steady$; this can be computed by measuring the sensitivities of  $\myav{R}_\steady$ and $\myav{S}_\steady$ to the production and degradation rates of the signal molecule. Let us suppose that  the signal $\myav{S}_\steady$ is varied by changing its production
rate $k_s$ infinitesimally at fixed degradation rate $k_d$ (we could
have chosen instead to vary $k_d$ or, in principle, both $k_s$ and
$k_d$). The gain is then
\begin{equation}
g=\frac{\partial\ln\myav{R}_\steady/\partial k_s}%
{\partial\ln\myav{S}_\steady/\partial k_s}\,=%
\frac{k_s}{\myav{R}_\steady}\frac{\partial \myav{R}_\steady}{\partial k_s}
\label{eq:gain}
\end{equation}
where the second equality follows from the fact that
${\partial\ln\myav{S}_\steady/\partial k_s}=1/k_s$, since
$\myav{S}_\steady=k_s/k_d$. We use the methods described in Section
\ref{sec:ss} to compute the steady-state sensitivity $\partial
\myav{R}_\steady/\partial k_s$, and hence the gain $g$.

Figure \ref{fig:stochf} shows the absolute differential gain $|g|$
computed using the time-averaged correlation function method, with
time-step pre-averaging, as a function of the signal strength
$\myav{S}_\steady$, as $k_s$ is varied (note that in this region the
actual gain is negative so $|g|=-g$).  The results are in excellent
agreement with the finite state projection method (FSP, see
Supplementary Material Section \ref{sec:fsp}).
Fig.~\ref{fig:stochf} (dashed line) also shows the mean-field theory
prediction derived from the second of Eqs.~\eqref{eq:stochfmft},
namely $g=\myav{S}_\steady/(\myav{S}_\steady+K)$. Stochastic focusing, as predicted by
Paulsson {\em{et al}} \cite{PBE00}, is clearly evident: the gain is
much greater in magnitude for the stochastic model than the mean-field
theory predicts, implying that fluctuations greatly increase the
sensitivity of the output signal to the input signal \cite{note-sf}.

In this example, the parameter of interest ($k_s$) is the rate
constant of a single reaction (production of S). As discussed in
Section \ref{sec:tw}, this implies that the parameter sensitivity can
be computed simply by counting the number of times this reaction is
visited, which can be achieved by modifying the reaction scheme to
\begin{equation}
\emptyset\overset{k_s}\to \mathrm{S} + \mathrm{Q}\,,\quad
\mathrm{S}\overset{k_d}\to\emptyset\,,\quad
\emptyset\overset{k^\star}\to \mathrm{R} \overset{1}\to\emptyset\,.
\end{equation}
then computing the weight function 
\begin{equation}
W_{k_s}=\frac{1}{k_s}\left[Q(t)-k_s t\right]\,.
\label{eq:qtrick}
\end{equation}
(which is the analogue of Eq.~\eqref{eq:key2c}), and using this to
obtain the relevant time-correlation functions. This requires no
changes to the simulation algorithm, making it easy to use with
existing software packages. As a demonstration, we computed the
differential gain for the parameters in Eq.~\eqref{eq:stochfpar},
using the open source simulation package \COPASI\ \cite{HS06}. To
achieve this, we used the Gibson-Bruck algorithm (as implemented in
\COPASI) to generate samples of $W_{k_s}$ and $R$ at equi-spaced
time points with a spacing $\Delta t=10$ time units (chosen to be
longer than the expected relaxation time of the output signal, set by
the decay constant for R). By taking the difference between successive
time points we compute $\Delta W_{k_s}$ and hence the
correlation function defined in Eq.~\eqref{eq:cdeftan}. The result,
shown in blue in Fig.~\ref{fig:stochf}, is in good agreement with our
other calculations.

\begin{figure}
\begin{center}
\includegraphics{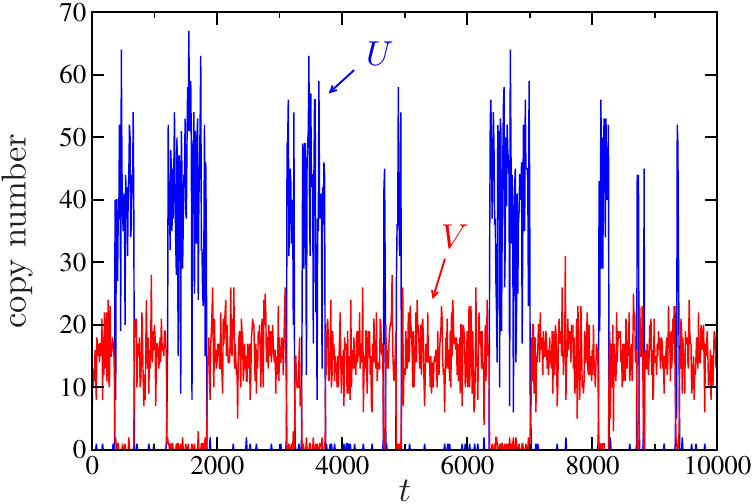}
\end{center}
\caption[?]{(color online) Representative time traces of $U$ and $V$
  in the Gardner \etal\ genetic switch model (Section
  \ref{sec:gardsw}).  Parameters are $\alpha_1=50$, $\beta=2.5$,
  $\alpha_2=16$, and $\gamma=1$.\label{fig:gardsw_cps}}
\end{figure}

\begin{figure}
\begin{center}
\includegraphics{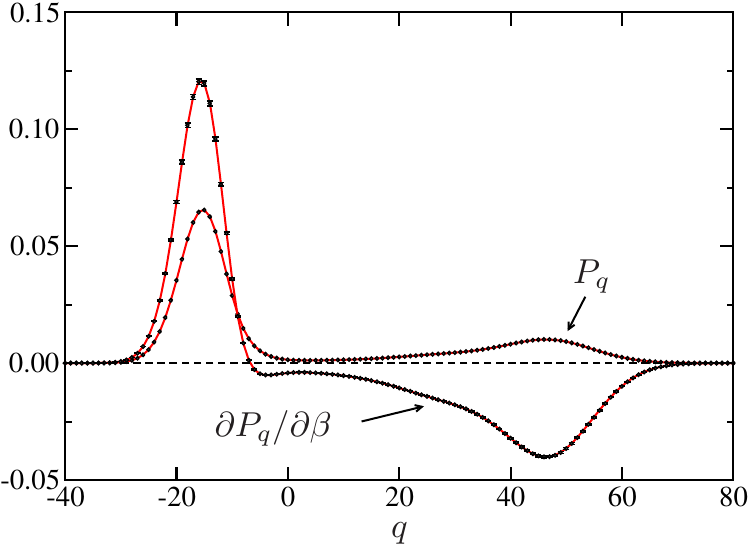}
\end{center}
\caption[?]{(color online) Steady state probability distribution for
  the order parameter $q\equiv U-V$ in the Gardner \etal\ switch, and
  the sensitivity to $\beta$.  Points (with small error bars in the
  case of the sensitivity) are from Gillespie simulations using the
  time-average correlation function method with time step
  pre-averaging; error bars are estimated by averaging over 10
  trajectories of $10^9$ steps. The history array length was
  $n=5\times10^4$.  The solid lines (red) are the results of FSP
  applied to this problem.  Model parameters are as in
  Fig.~\ref{fig:gardsw_cps}.\label{fig:gardsw_op}}
\end{figure}

\subsection{A bistable genetic switch}
\label{sec:gardsw}
As a final example, we consider a model for a bistable genetic switch,
of the type constructed experimentally by Gardner \etal\ \cite{GCC00},
in which two proteins U and V mutually repress each other's
production. We suppose that transcription factor binding to the
operator is cooperative, and can be described by a Hill function; the
rate of production of protein U is then given by
$\alpha_1/(1+V^\beta)$ while the rate of production of V is given by
$\alpha_2/(1+U^\gamma)$ (here $\alpha_1$ and $\alpha_2$ describe the
maximal production rates while $\beta$ and $\gamma$ are the Hill
exponents, describing the degree of cooperativity). The units of time
are fixed by setting the degradation rates of U and V to unity.  For a
suitable choice of parameter values, stochastic simulations of this
model show switching between a U-rich state and a V-rich state, as
illustrated in Fig.~\ref{fig:gardsw_cps}; the steady-state probability
distribution $P_q$ for the quantity $q \equiv U-V$ is bimodal, as
shown in Fig.~\ref{fig:gardsw_op}. We use this example to illustrate
the computation of parameter sensitivities for more complicated
scenarios where the system property of interest is not simply a mean
copy number and the parameter of interest is not a simple rate
constant. In particular we compute the sensitivity of the steady-state
probability distribution $P_q$ to the Hill exponent $\beta$.

Our model consists of the following reaction scheme:
\begin{equation}
{}\overset{1}\to \mathrm{U} \overset{2}\to{},\qquad
{}\overset{3}\to \mathrm{V} \overset{4}\to{}.\qquad
\end{equation}
in which proteins U and V are created and destroyed with propensities given by:
\begin{equation}
a_1=\frac{\alpha_1}{1+V^\beta},\quad a_2=U,\quad
a_3=\frac{\alpha_2}{1+U^\gamma},\quad a_4=V
\label{eq:gardprop}
\end{equation}
Let us first suppose we wish to compute $\partial P_q/\partial \beta$
using the time-averaged correlation function method, without time step
pre-averaging.  We use the propensity functions in
Eqs.~\eqref{eq:gardprop} to run a standard kinetic Monte Carlo
(Gillespie) simulation, choosing at each step a next reaction and a
time step $\delta t$. At each simulation step, we also compute the
quantity
\begin{equation}
z\equiv\frac{\partial\ln a_1}{\partial\beta}=-\frac{V^\beta \ln
  V}{1+V^\beta}
\label{eq:zed}
\end{equation}
and update the weight function $W$ according to Eq.~\eqref{eq:key2b},
\ie\ if reaction 1 is chosen as the next reaction, we increment
$W$ by $z(1-a_1 \delta t)$, otherwise, we increment $W$ by $-z a_1
\delta t$ (note that it is correctly $a_1$ that features in this,
irrespective of the chosen next reaction). We keep track not only of
the current value of $W$, but also of its value a fixed number $n$
steps ago.  At the same time, we keep track of the function of
interest (denoted $f$ in Sections \ref{sec:tw} and
\ref{sec:ss}). Because we are computing the parameter sensitivity of
the {\em{distribution}} $P_q$, we have a function $f_q$, and a
time-correlation function $C_{q,\mathrm{time\,\,av}}(n)$, for
{\em{each}} value of $q$. At each simulation step, we check the
current value of $q(t)$. The function $f_q$ is unity if $q=q(t)$ and
zero otherwise (\ie\  $f_q = \delta_{q(t),q}$). For each value
of $q$, we then compute the time correlation function
$C_{q,\mathrm{time\,\,av}}(n)$ as prescribed in
Eq.~\eqref{eq:cdeftan}. As long as $n \bar{\tau}$ (where $\bar{\tau}$
is the global average time step) is longer than the typical relaxation
time of the system, $C_{q,\mathrm{time\,\,av}}(n)$ should give a good
estimate for $\partial P_q/\partial \beta$.

If we are instead using time step pre-averaging, we employ a slight
modification of the above procedure. At each simulation step, we
choose a next reaction, but we do not choose a time step $\delta
t$. In our update rules for $W$, we replace $\delta t$ by the
state-dependent mean timestep $\tau$ where $\tau^{-1}=\sum_{\mu=1}^4
a_\mu$. As well as keeping track of $W$ and $f_q$ we also need to
compute at each step
\begin{equation}
\frac{\partial\tau}{\partial\beta}=
a_1\frac{\partial\tau}{\partial a_1}\cdot
\frac{\partial\ln a_1}{\partial\beta}=
-\tau^2a_1z\,.
\end{equation}
This quantity is then used to compute $C_{q,\mathrm{time\,\,av}}(n)$
and hence $\partial P_q/\partial \beta$ using the modified algorithm
given in Supplementary Material Section \ref{sec:preav}.

An important technical point here concerns the relaxation time of the
system, or the number of steps $n$ over which we need to remember the
system's history in order that the correlation function $C(n)$ gives a
good estimate of the steady-state parameter sensitivity. For the
previous examples studied, this timescale was given by the slowest
decay rate (typically that of the protein molecules). The genetic
switch, however, shows dynamical switching behaviour on a timescale
that is much longer than the protein decay rate (see for example
Fig.~\ref{fig:gardsw_cps}). We therefore need to choose a value of $n$
such that $n \bar{\tau}$ is longer than the typical switching time.
Kinetic Monte-Carlo simulations (like those in
Fig.~\ref{fig:gardsw_cps}) show that for our model, the typical time
between switching events is approximately 160 time units, while the
global average time step $\bar{\tau} \approx 0.02$. The typical number
of steps per switching event is therefore $160/0.02 = 8 \times 10^3$.
Our chosen value of $n$ should be at least this large. In practice we
find that the correlation functions are fully converged (to within a
reasonable accuracy) by $n=5\times 10^4$ steps ($\approx6.3$ switching
events), but not quite converged by $n=2\times 10^4$ steps
($\approx2.5$ switching events).  These lengthy convergence times mean
that much longer simulations are needed to obtain good statistical
estimates for the parameter sensitivity in this model than in the
previous examples.

Figure \ref{fig:gardsw_op} shows the steady state probability
distribution $P_q$ together with its sensitivity
$\partial P_q/\partial \beta$, computed using the time-averaged
correlation function method with time step pre-averaging, for the same
parameters as in Fig.~\ref{fig:gardsw_cps}. This method gives results
in excellent agreement with FSP.  $P_q$ has the bimodal shape typical
of a stochastic genetic switch, with a large peak at $q\approx-15$ and
a much broader peak around $q\approx45$, with a minimum around
$q\approx5$.  The sensitivity coefficient $\partial P_q/\partial\beta$
measures how the behaviour of the switch depends on the cooperativity
$\beta$ of binding of the transcription factor V.  We see that
increasing $\beta$ leads to an increased peak at $q<0$, and a
decreased peak at $q>0$, in other words the switch spends more time in
the V-rich state.  Also the minimum around $q\approx5$ decreases,
suggesting that the switching frequency decreases as $\beta$
increases. This is confirmed by further study using the
ensemble-averaged correlation function method of the sensitivity
coefficient of the switching frequency to changes in $\beta$; the
details of this will be presented elsewhere.

\section{Discussion}
In this paper, we have shown how trajectory reweighting can be used to
compute parameter sensitivity coefficients in stochastic simulations
without the need for repeated simulations with perturbed values of the
parameters. The methods presented here are simple to implement in
standard kinetic Monte Carlo (Gillespie) simulation algorithms and in
some cases can be used without any changes to the simulation code,
making them compatible with packages such as \COPASI\ \cite{HS06}. For
computation of time-dependent sensitivity coefficients, the method
involves tracking a weight function (which depends on the derivative
of the propensities with respect to the parameter of interest) and
computing its covariance with the system property of interest, at the
time of interest, across multiple simulations. For computing
time-independent steady-state parameter sensitivities, we show that
the sensitivity coefficient can be obtained as the long-time limit of
a time correlation function, which can be computed either across
multiple simulations (ensemble-averaged correlation function method),
or as a time average in a single simulation run (time-averaged
correlation function method). We further show that time step
pre-averaging removes the need to choose a new time step at each
simulation step, significantly improving computational efficiency. In either the time-dependent or the time-independent case, it is a trivial matter to compute  multiple sensitivity coefficients (e.g. with respect to different parameters) at the same time -- one simply tracks each of the corresponding weight functions simultaneously.

In deterministic models, parameter sensitivity coefficients can be
computed by simultaneous integration of a set of {\em{adjunct}} ODEs,
alongside the set of ODEs describing the model (see
Eq.~\eqref{eq:1}). We consider the trajectory reweighting approach
described here to be the exact stochastic analogue of the adjunct ODE
method; the integration of the adjunct ODEs alongside the original
ODEs is directly analogous to the procedure of generating a trajectory
weight alongside the normal trajectory in a kinetic Monte-Carlo
scheme.  Indeed, one can derive an {\em{adjunct chemical master
    equation}} by taking the derivative of the chemical master
equation with respect to the parameter of interest; it turns out that
the trajectory reweighting scheme is essentially a stochastic solution
method for the adjunct master equation \cite{AW-2bpub}.

In Section \ref{sec:stochf}, we demonstrated the use of trajectory
reweighting to compute parameter sensitivities, and hence the differential gain, for a model of a
stochastic signaling network. We believe that this approach has
widespread potential application to signaling pathways, because it can
be implemented for any existing model without any modifications to the
underlying kinetic Monte-Carlo simulation code. As long as a
stochastic input signal is generated by a process
$\emptyset\to\mathrm{S}\to\emptyset$, one can use the ghost particle
trick to compute the sensitivity of any quantity of interest to the
input signal (controlled by varying the rate $k_s$ of the signal
production reaction) by modifying the production step to
$\emptyset\to\mathrm{S}+\mathrm{Q}$, computing the weight function
from $W_{k_s}=Q/k_s-t$ (see Eq.~\eqref{eq:qtrick}) and computing the
appropriate correlation function for its covariance with the system
property of interest.  As proof-of-principle we calculated the
differential gain for the model in Section \ref{sec:stochf} (see
Fig.~\ref{fig:stochf}), using \COPASI\ \cite{HS06} to generate the
simulation data, and a standard spreadsheet package to compute the
correlation function.

In Section \ref{sec:gardsw}, we used the methodology to compute the
sensitivity of the probability distribution function for a bistable
genetic switch, to the degree of cooperativity (Hill exponent) of
binding of one of its transcription factors. This example demonstrates
that trajectory reweighting is not a panacea for all problems. The
bistable genetic switch has a long relaxation time, which requires the
correlation function of the weight to the computed over long times,
with a corresponding need for large sample sizes to obtain good
statistical sampling. While trajectory reweighting works for this
example, preliminary attempts to compute the parameter dependence of
the switching rate show that finite differencing may be more
efficient.  In fact Plyasunov and Arkin \cite{PA07} already discuss in
which cases it may be more efficient to use
finite-differencing. Because of their long relaxation times, genetic
switches are notoriously difficult to study in stochastic simulations
\cite{MtW09}.  A plethora of sophisticated schemes have been developed to
address this problem \cite{AWtW05,Allen09,DD10}, some of which could perhaps
be extended to incorporate trajectory reweighting.

The present study considers how to compute parameter sensitivity
coefficients---\ie\ first derivatives of system properties with
respect to the parameters. The same approach can, however, easily be
used to compute higher derivatives, such as the Hessian matrix, as
discussed in Supplementary Material Section \ref{sec:rat}. This raises
the possibility of combining the present methods with gradient-based
search algorithms, to make a sophisticated \emph{parameter estimation}
algorithm for stochastic modeling.  This would offer a novel approach
to a major class of problems in systems biology.

To summarise, we believe the trajectory reweighting schemes presented
here are an important and useful addition to the stochastic simulation
toolbox.  Further research should address in detail their performance
with respect to existing methods \cite{KDN07, PA07,RSK10} and their
application to challenging models such as those with long relaxation
times, as well as their potential for use in more sophisticated
parameter search algorithms.

\begin{acknowledgments}
The authors thank Mustafa Khammash for detailed advice about the FSP
method and assistance with its implementation. RJA was supported by a
Royal Society University Research Fellowship.  The collaboration
leading to this work was facilitated by the StoMP research network
under BBSRC grant BB/F00379X/1 and by the e-Science Institute under
theme 14: ``Modelling and Microbiology''.
\end{acknowledgments}

\begin{thebibliography}{23}
\bibitem[{not({\natexlab{a}})}]{note-error}
\bibinfo{note}{The fact that the two terms in Eq.~\eqref{eq:key2a} cancel on
  average could in principle be used to simplify Eq.~\eqref{eq:key1} to
  ${\partial\myav{f}}/{\partial k_\alpha}=\myav{f W_{k_\alpha}}$. However, in
  practice we have found that retaining the term in $\myav{W_{k_\alpha}}$
  improves the statistical sampling (for example if $f$ is a constant then the
  right hand side of Eq.~\eqref{eq:key1} vanishes exactly, as it should, even
  if the sampling error causes $\myav{W_{k_\alpha}}\ne0$).}
\bibitem[{not({\natexlab{d}})}]{note-epar}
\bibinfo{note}{If this is not the case one can always make it so by
augmenting the parameter set.}
\bibitem[{not({\natexlab{b}})}]{note-efficiency} \bibinfo{note}{The
  efficiency gain obtained by time-step pre-averaging depends of
  course on the balance between computing time steps and choosing
  reactions: this might differ for different algorithms and reaction
  schemes.}
\end{thebibliography}


\begin{widetext}

\vspace{12pt}

\begin{center}
\framebox{\bf\Large SUPPLEMENTARY MATERIAL}
\end{center}


\setcounter{section}{0}
\def\thesection{\arabic{section}}

\large

\section{Derivatives of averages with respect to parameters}
\label{sec:rat}
Here, we present a convenient way to compute the derivatives of
average quantities with respect to the parameters of the model, that
are required to arrive at Eqs.~\eqref{eq:key1} and
\eqref{eq:key2} in the main text. We also show that this method
generalizes easily to higher derivatives.

Noting that in the perturbed system the parameter $k_\alpha$ has been
changed to $k_{\alpha}'$, we use Eq.~\eqref{eq:pppp} in the main
text to write the average of the function $f$ in the perturbed system
as
\begin{equation}
\myav{f}'=
\frac{\myav{f R(k'_\alpha,k_\alpha)}}
{\myav{R(k'_\alpha,k_\alpha)}}
\label{eq:ppp2}
\end{equation}
where
\begin{equation}
R(k'_\alpha,k_\alpha)\equiv \frac{P(k'_\alpha)}{P(k_\alpha)}
=e^{\ln P(k'_\alpha)-\ln P(k_\alpha)}\,.
\end{equation}
The function $R(k'_\alpha,k_\alpha)$ has the property that ${\partial R}/{\partial
  k'_\alpha}=W'_{k'_\alpha}R$ where $W'_{k'_\alpha}={\partial\ln
  P(k'_\alpha)}/{\partial k'_\alpha}$.  We then have
\begin{equation}
\frac{\partial \myav{f}'}{\partial k'_\alpha}=
\frac{\myav{f W'_{k'_\alpha} R}}
{\myav{ R}}-
\frac{\myav{f R}
\myav{W'_{k'_\alpha} R}}
{\myav{R}^2}
\,.
\end{equation}
Taking the limit $k'_\alpha\to k_\alpha$ (for an infinitesimal
perturbation), and noting thereby that $R\to1$ and $W'_{k'_\alpha}\to
W_{k_\alpha}$, yields Eqs.~\eqref{eq:key1} and
\eqref{eq:key2} in the main text.

Taking this procedure further allows the computation of higher
derivatives; one can show for instance that the Hessian is
\begin{equation}
\begin{array}{l}
\displaystyle
\frac{\partial^2\myav{f}}{\partial k_\alpha\partial k_\beta}
=\myav{f W_{k_\alpha}  W_{k_\beta}} +\myav{f W_{k_\alpha k_\beta}}
- \myav{f W_{k_\alpha}}\myav{W_{k_\beta}}
- \myav{f W_{k_\beta}}\myav{W_{k_\alpha}}\\[6pt]
{}\hspace{10em}
- \myav{f}\myav{ W_{k_\alpha}W_{k_\beta}}
- \myav{f}\myav{ W_{k_\alpha k_\beta}}
+2\myav{f}\myav{ W_{k_\alpha}}\myav{W_{k_\beta}}\,.
\end{array}
\label{eq:fab}
\end{equation}
where
\begin{equation}
W_{k_\alpha k_\beta}=\frac{\partial^2\!\ln P}
{\partial k_\alpha\partial k_\beta}\,.
\end{equation}
Eq.~\eqref{eq:fab} is potentially useful for gradient search
algorithms.  This expression is likely to simplify in many cases --
for instance we expect that ${\partial^2\!\ln P}/{\partial
  k_\alpha\partial k_\beta}$ often vanishes for $k_\alpha\ne k_\beta$.
One might also use the fact that
$\myav{W_{k_\alpha}}=\myav{W_{k_\beta}}=0$, but it may improve the
statistical sampling to retain these terms (see discussion in main
text).

\section{Exact results for problems with linear propensities}
\label{sec:linear}
In this Section we describe some exact results that can be obtained
for models with linear propensity functions, in particular for the
correlation functions defined in Eqs.~\eqref{eq:cdef} and
\eqref{eq:DWt} in the main text.  The analysis draws heavily on
established literature results (which we summarize below).  More
details and links to earlier literature can be found in the appendix
to Supplementary Ref.~\onlinecite{WTtW06}.

To fix notation, let us suppose that the $\alpha$-th propensity
function depends linearly on the copy numbers $N_j$, namely
$a_\alpha=\sum_j K_{\alpha j}N_j+b_\alpha$
where  $K_{\alpha
  j}$ and $b_\alpha$ are constants which we assume to be proportional to
the rate consant $k_\alpha$. Our aim is to compute the sensitivity coefficients
$\partial\myav{N_i}/\partial\ln k_\alpha$.

It is well known that for linear propensity functions the moment equations
close successively.  Thus, the mean copy numbers $\myav{N_i}$ obey
\begin{equation}
\frac{\partial\myav{N_i}}{\partial t}={\textstyle\sum_\alpha} 
\nu_{i\alpha}\myav{a_\alpha}
={\textstyle\sum_j} K_{ij}\myav{N_j}+b_i
\label{eq:ni}
\end{equation}
where $\nu_{i\alpha}$ is the stoichiometry matrix (describing the change
in the copy number of the $i$-th species due to the firing of the
$\alpha$-th reaction),  $K_{ij}=\sum_\alpha
\nu_{i\alpha}K_{\alpha j}$ and $b_i=\sum_\alpha
\nu_{i\alpha}b_\alpha$.  Note that $K_{ij}$ is usually asymmetric.
For the second moments, the variance-covariance matrix
$S_{ij}(t)=\myav{\Delta N_i(t)\,\Delta N_j(t)}$,
where $\Delta N_i=N_i-\myav{N_i}$, obeys
\begin{equation}
\frac{\partial S_{ij}}{\partial t}={\textstyle\sum_k}
(K_{ik}S_{jk}+K_{jk}S_{ik})+H_{ij}
\label{eq:sij}
\end{equation}
where
\begin{equation}
H_{ij}={\textstyle\sum_\alpha}\nu_{i\alpha}\nu_{j\alpha}\myav{a_\alpha}.
\label{eq:hij}
\end{equation}
Note that $S_{ij}(t)$ is symmetric.
Finally the time-ordered two-point correlation
functions $\myav{\Delta N_i(t)\,\Delta N_j(t')}$ with $t>t'$ obey a
regression theorem
\begin{equation}
\frac{\partial \myav{\Delta N_i(t)\,\Delta N_j(t')}}{\partial t}
={\textstyle\sum_k}K_{ik}
\myav{\Delta N_k(t)\,\Delta N_j(t')}.
\label{eq:regr}
\end{equation}
This concludes our survey of the established literature results.

\begin{figure}
\begin{center}
\includegraphics{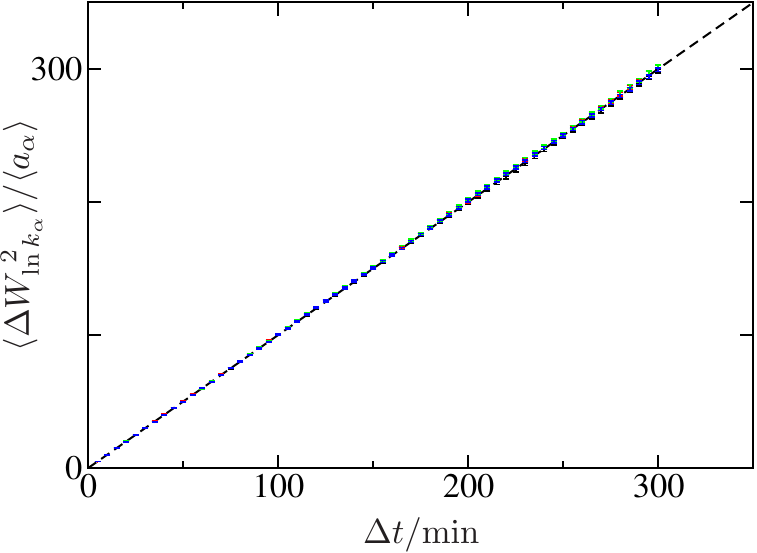}
\end{center}
\caption[?]{(color online) Growth of variance of weight functions in
  steady state, normalised by the expected growth rate.  Model and
  parameters as for Fig.~\ref{fig:corrn} in the main
  text.\label{fig:dwsq}}
\end{figure}

We now employ the ghost particle trick of Section \ref{sec:tw} in
the main text, and suppose that reaction $\alpha$ now creates a
noninteracting species $Q_\alpha$, in addition to its usual
products. Following Eq.~\eqref{eq:ni}, the mean copy number
$\myav{Q_\alpha}$ obeys
\begin{equation}
\frac{\partial\myav{Q_\alpha}}{\partial t}=\myav{a_\alpha}.
\label{eq:qa}
\end{equation}
For the second moments,
Eq.~\eqref{eq:sij} becomes
\begin{equation}
\frac{\partial S_{i\alpha}}{\partial t}={\textstyle\sum_j}
(K_{ij}S_{j\alpha}+K_{\alpha j}S_{ij})+H_{i\alpha}.
\label{eq:i2}
\end{equation}
where $S_{i\alpha}=\myav{\Delta N_i\,\Delta Q_\alpha}$.
The second term in this can be rewritten as
\begin{equation}
{\textstyle\sum_j}
K_{\alpha j}S_{ij}=
\myav{\Delta N_i({\textstyle\sum_j}K_{\alpha j}\Delta N_j)}
=\myav{\Delta N_i\Delta a_\alpha}.
\end{equation}
The stoichiometry matrix entry for $Q_\alpha$ consists of a `$+1$' for
the $\alpha$-th reaction, and zero elsewhere, so that Eq.~\eqref{eq:hij} becomes
$H_{i\alpha}=\nu_{i\alpha}\myav{a_\alpha}$. Finally we have
\begin{equation}
\frac{\partial S_{i\alpha}}{\partial t}={\textstyle\sum_j}
K_{ij}S_{j\alpha}+\myav{\Delta N_i\Delta a_\alpha}
+\nu_{i\alpha}\myav{a_\alpha}.
\label{eq:sia}
\end{equation}
By similar argumentation we also obtain
\begin{equation}
\frac{\partial S_{\alpha\alpha}}{\partial t}
=2\myav{\Delta Q_\alpha\Delta a_\alpha}+\myav{a_\alpha}
\label{eq:saa}
\end{equation}
where $S_{\alpha\alpha}=\myav{\Delta Q^2_\alpha}$.

Armed with these results, let us now turn to the problem of computing
the weight function $W_{k_\alpha}$.  We write the continuous-time
analogue of Eq.~\eqref{eq:key2c} in the main text:
%
\begin{equation}
k_\alpha W_{k_\alpha}(t)=%
Q_\alpha(t)-\int_0^t \!dt'\, a_\alpha(t')
\label{eq:qa2}
\end{equation}
(note that by substituting Eq.~\eqref{eq:qa} into
Eq.~\eqref{eq:qa2} we recover our previous observation that
$\myav{W_{k_\alpha}}=0$).  Again writing $W_{\ln k_\alpha} \equiv
k_\alpha W_{k_\alpha}$, it follows from
Eq.~\eqref{eq:qa2} that
\begin{equation}
\myav{N_i W_{\ln k_\alpha}} =\myav{\Delta N_i \Delta W_{\ln k_\alpha}}
= S_{i\alpha}(t)
-\int_0^t \!dt'\, \myav{\Delta N_i(t)\,\Delta a_\alpha(t')}.
\label{eq:nw}
\end{equation}
Thus, now noting explicitly the time-dependence of the various terms, we obtain
\begin{equation}
\frac{\partial \myav{N_i W_{\ln k_\alpha}}}{\partial t}=
\frac{\partial S_{i\alpha}}{\partial t}
-\myav{\Delta N_i(t)\,\Delta a_\alpha(t)}
-{\int_0^t \!\!dt'}\, \frac{\partial\myav{\Delta N_i(t)\,\Delta
    a_\alpha(t')}}{\partial t}.
\end{equation}
Exploiting the linearity of the propensity functions, the
regression theorem implies
\begin{equation}
\frac{\partial\myav{\Delta N_i(t)\,\Delta
    a_\alpha(t')}}{\partial t}
={\textstyle\sum_j} 
K_{ij}\myav{\Delta N_j(t)\,\Delta a_\alpha(t')}.
\end{equation}
Hence
\begin{equation}
\frac{\partial \myav{N_i W_{\ln k_\alpha}}}{\partial t}=
\frac{\partial S_{i\alpha}}{\partial t}
-\myav{\Delta N_i(t)\,\Delta a_\alpha(t)}
-\sum_j K_{ij}\int_0^t \!\!dt'\, \myav{\Delta N_j(t)\,\Delta
    a_\alpha(t')}.
\label{eq:i1}
\end{equation}
Eliminating the time integrals between this and Eq.~\eqref{eq:nw}, we get
\begin{equation}
\frac{\partial \myav{N_i W_{\ln k_\alpha}}}{\partial t}=
\frac{\partial S_{i\alpha}}{\partial t}
-\myav{\Delta N_i(t)\,\Delta a_\alpha(t)}
-\sum_j K_{ij} S_{j\alpha}(t)+\sum_j K_{ij} \myav{N_j W_{\ln k_\alpha}}.
\end{equation}
Eliminating $\partial S_{i\alpha}/\partial t$ between this and 
Eq.~\eqref{eq:sia} gives finally
\begin{equation}
\frac{\partial \myav{N_i W_{\ln k_\alpha}}}{\partial t}=
{\textstyle\sum_j} K_{ij} \myav{N_j W_{\ln k_\alpha}}
+\nu_{i\alpha}\myav{a_\alpha}.
\label{eq:a1}
\end{equation}
This is an ODE which give the evolution of $\myav{N_i W_{\ln
    k_\alpha}}$ in terms of known quantities.  It can be compared with
the adjunct ODE that is obtained by differentiating
Eq.~\eqref{eq:ni} with respect to $\ln k_\alpha$. The two ODEs are
identical and share the same initial conditions.  For this specific
case, this is a direct proof of the general result in the main text,
namely that $\myav{N_i W_{\ln k_\alpha}} \equiv {\partial
  \myav{N_i}}/{\partial\ln k_\alpha}$.  Whilst this is interesting, it
is not quite what we are after, which is a theory for the correlation
functions defined in the main text.  To find this we first note a
generalisation of the regression theorem is
\begin{equation}
\frac{\partial \myav{N_i(t)\,W_{\ln k_\alpha}(t_0)}}{\partial t}
={\textstyle\sum_k}K_{ij}
\myav{N_j(t)\,W_{\ln k_\alpha}(t_0)}\,.
\label{eq:a2}
\end{equation}
Subtracting this from Eq.~\eqref{eq:a1}
generates a set of coupled ODEs for the correlation functions
$C_i(t,t_0)=\myav{N_i(t)\Delta W_{\ln k_\alpha}(t,t_0)}$, which should
be solved with the initial conditions $C_i(t,t_0)=0$ at $t=t_0$.  In
steady state these ODEs are
\begin{equation}
\frac{\partial C_i(\Delta t)}{\partial(\Delta t)}=
{\textstyle\sum_j} K_{ij} C_j(\Delta t)
+\nu_{i\alpha}\myav{a_\alpha}\,.
\label{eq:a4}
\end{equation}
The initial conditions are $C_i(0)=0$.  This is the key result of this
Section, as in principle it allows for explicit calculation of the
correlation functions.  Comparing to the adjunct ODE obtained by
differentiating Eq.~\eqref{eq:ni} with respect to $\ln k_\alpha$,
we see that for this case we also have a direct proof of the general
result claimed in the main text, that $C_i(\Delta t)\to {\partial
  \myav{N_i}/}{\partial\ln k_\alpha}$ as $\Delta t\to\infty$.  For the
constitutive gene expression model in Section \ref{sec:santmac}
in the main text, we solved Eq.~\eqref{eq:a4} to obtain the
results given in Eq.~\eqref{eq:santmaccf} in the main text.

To complete the general discussion here, let us derive an expression
for the variance of $\Delta W_{\ln k_\alpha}$.  From the definition in
Eq.~\eqref{eq:qa2} we have
\begin{equation}
\myav{W^{{}\;2}_{\ln k_\alpha}} = S_{\alpha\alpha}-2\!\int_0^t \!\!dt'\,
\myav{\Delta Q_\alpha(t)\,\Delta a_\alpha(t')}
+\int_0^t \!\!dt'\int_0^t \!\!dt''\,
\myav{\Delta a_\alpha(t')\,\Delta a_\alpha(t'')}
\end{equation}
Thus
\begin{equation}
\frac{\partial \myav{W^{{}\;2}_{\ln k_\alpha}}}{\partial t} 
= \frac{\partial S_{\alpha\alpha}}{\partial t}
-2\myav{\Delta Q_\alpha\, \Delta a_\alpha}
-2\!\!\int_0^t \!\!dt'\,
\frac{\partial\myav{\Delta Q_\alpha(t)\,\Delta a_\alpha(t')}}{\partial t}
+2\!\!\int_0^t \!\!dt'\,
\myav{\Delta a_\alpha(t)\,\Delta a_\alpha(t')}.
\label{eq:dw2dt}
\end{equation}
The last two terms in this cancel, on account of the regression
theorem.  Further cancellations occur when Eq.~\eqref{eq:saa} for
$\partial S_{\alpha\alpha}/\partial t$ is inserted, giving finally
${\partial \myav{W^{{}\;2}_{\ln k_\alpha}}}/{\partial t}
=\myav{a_\alpha}$.  Since $W_{\ln k_\alpha}(t)=W_{\ln
  k_\alpha}(t_0)+\Delta W_{\ln k_\alpha}$, and $W_{\ln k_\alpha}(t_0)$
and $\Delta W_{\ln k_\alpha}$ are uncorrelated, it follows that
$\myav{W^{{}\;2}_{\ln k_\alpha}(t)} =\myav{W^{{}\;2}_{\ln
    k_\alpha}(t_0)} +\myav{\Delta W^{{}\;2}_{\ln k_\alpha}}$.
Integrating ${\partial \myav{W^{{}\;2}_{\ln k_\alpha}}}/{\partial t}
=\myav{a_\alpha}$ and inserting in this last expression gives
$\myav{\Delta W^{{}\;2}_{\ln k_\alpha}}
={\int_{t_0}^t}\!dt'\,\myav{a_\alpha}$.  As a particular case, in
steady state, $\myav{\Delta W^{{}\;2}_{\ln k_\alpha}}
=\myav{a_\alpha}\,\Delta t$.  Thus we do indeed see that in steady
state $\Delta W_{\ln k_\alpha}$ has a zero mean and a variance that
grows linearly in time, justifying our claim that it behaves
essentially like a random walk.  Some results confirming this analysis
are shown in Fig.~\ref{fig:dwsq}.

\section{Time step pre-averaging}
\label{sec:preav}
In the time step pre-averaging approach, we do not select time steps
as part of our kinetic Monte Carlo algorithm, but instead use the
state-dependent average time step $\tau\equiv(\sum_\mu a_\mu)^{-1}$ in
our expression for the time average of quantity $f$, as in
Eq.~\eqref{eq:pre} in the main text. For the purposes of this
Section, we define a new notation:
\begin{equation}
\mynav{f\tau}=\frac{\sum_{\mathrm{steps}}\,f\tau}{N_{\mathrm{steps}}}
\label{eq:avdef}
\end{equation}
in which the sum is over the values of the system function $f$
multiplied by the (state-dependent) mean time step, computed at each
step along a kinetic Monte-Carlo trajectory of length
$N_{\mathrm{steps}}$.  Note that $\mynav{f\tau}$ can be computed using
an algorithm that does not keep track of time but only of the choice
of reaction channel. We can then rewrite Eq.~\eqref{eq:pre} in the
main text as
\begin{equation}
\overline {f} =
\frac{\sum_{\mathrm{steps}}\,f\tau}{\sum_{\mathrm{steps}}\,\tau} %
= \frac{\mynav{f\tau}}{\mynav{\tau}}\,.
\label{eq:pre2}
\end{equation}  

When using time step pre-averaging, the correlation function
Eq.~\eqref{eq:cdeftan} in the main text must be modified because
the relative probability of generating a given sequence of states
(Eq.~\eqref{eq:Pdef} in the main text) takes a different form
when the algorithm does not keep track of time, and because the
average time step $\tau$ in Eq.~\eqref{eq:pre2} itself usually depends
on the parameter in question.

In a kinetic Monte Carlo scheme in which the next reaction is selected
as normal, but time is not tracked, the probability of generating a
given trajectory is proportional to
\begin{equation}
P=\textstyle{\prod_{\mathrm{steps}}}\,
(a_\mu/\sum_\mu a_\mu).
\label{eq:nPdef}
\end{equation}
(i.e. the part of Eq.~\eqref{eq:Pdef} in the main text concerning
the time step distribution is discarded). In analogy to
Eq.~\eqref{eq:pppp} in the main text, one can write the average
$\mynavprime{f\tau}$ for the perturbed problem in terms of averages
over unperturbed trajectories:
\begin{equation}
\mynavprime{f\tau}=\frac{\mynav{f\tau' P'/P}}{\mynav{P'/P}}\,.
\label{eq:avepet}
\end{equation}
Taking derivatives of Eq.~\eqref{eq:avepet} with respect to the
parameter $k_\alpha$ as described in Section \ref{sec:rat} above, it
follows that
\begin{equation}
\label{eq:www}
\frac{\partial\mynav{f\tau}}{\partial k_\alpha}
=\mynav{f\frac{\partial\tau}{\partial k_\alpha}}
+\mynav{f\tau W_{k_\alpha}}-\mynav{f\tau}\mynav{W_{k_\alpha}}
\end{equation}
where the fact that $\tau$ depends on $k_\alpha$ leads to an extra
term (the first term) in comparison to Eq.~\eqref{eq:key1} in the
main text.  Computing $W_{k_\alpha}=\partial\ln P/\partial k_\alpha$
using Eq.~\eqref{eq:nPdef}, it turns out that $W_{k_\alpha} =
\sum_{\mathrm{steps}}\,\delta W_{k_\alpha}$ has the same form as
before, but with $\delta t$ replaced by $\tau$,
\begin{equation}
\delta W_{k_\alpha}=\frac{\partial\ln a_\mu}{\partial k_\alpha}
-\frac{\partial(\sum a_\mu)}{\partial k_\alpha}\tau.
\label{eq:dww}
\end{equation}
Since the weight function in Eq.~\eqref{eq:dww} behaves like a random
walk, steady-state parameter sensitivities should be computed using
the correlation function trick (as in Section \ref{sec:ss} in the
main text).  From Eq.~\eqref{eq:www} we have:
\begin{equation}
\frac{\partial\mynav{f\tau}}{\partial k_\alpha}
=\mynav{f\frac{\partial\tau}{\partial k_\alpha}}
+\lim_{n\to\infty}C(n)
\label{eq:key42a}
\end{equation}
where
\begin{equation}
C(n)=\mynav{f\tau \Delta W_{k_\alpha}(n)}
-\mynav{f\tau}\mynav{\Delta W_{k_\alpha}(n)}
\label{eq:key42b}
\end{equation}
with $\Delta W_{k_\alpha}(n)$ given by Eq.~\eqref{eq:DWn} in the
main text, using the present Eq.~\eqref{eq:dww} to generate
$W_{k_\alpha}$. Finally, the parameter sensitivity coefficient itself
is given by differentiating Eq.~\eqref{eq:pre2},
\begin{equation}
\frac{\partial \overline{f}}{\partial k_\alpha}=
\frac{1}{\mynav{\tau}}\frac{\partial\mynav{f\tau}}{\partial k_\alpha}
-\frac{\mynav{f\tau}}{\mynav{\tau}^2}
\frac{\partial\mynav{\tau}}{\partial k_\alpha}\,.
\label{eq:tspre}
\end{equation}
The quantity ${\partial\mynav{\tau}}/{\partial k_\alpha}$ in the
second term is given by setting $f=1$ in Eqs.~\eqref{eq:key42a} and
\eqref{eq:key42b}.  When using time step pre-averaging in combination
with the time-averaged correlation function method, one computes the
parameter sensitivity coefficient using Eq.~\eqref{eq:tspre} rather
than simply taking the limit of the correlation function as $n \to
\infty$.

We note that while Eq.~\eqref{eq:tspre} looks formidable, its actual
computation is fairly straightforward.  To obtain both $\overline{f}$
and $\partial\overline{f}/{\partial k_\alpha}$, one computes
trajectory averages of the set of quantities defined by $\{1,
f\}\otimes \{1, \tau\} \otimes \{1, \Delta W_{k_\alpha}(n)\}$,
together with $\partial\tau/\partial k_\alpha$ and $f
\partial\tau/\partial k_\alpha$.  These averages are calculated by
summing the respective quantities along the trajectory and dividing by
the number of steps.

\section{The finite state projection algorithm}
\label{sec:fsp}
The master equation describes the evolution of the probability $P(N_i;
t)$ that a system is in the state $N_i$ at time $t$.  For the sake of
compactness we will adopt the notation $s$ and $s'$ for the states
$N_i$ and $N_i'$ respectively.  The master equation is \cite{vK92}
\begin{equation}
\frac{\partial P_s}{\partial t}
={\textstyle\sum_{s'}}[w_{s'\!s}P_{s'}-w_{ss'}P_s]
\label{eq:master}
\end{equation}
where $w_{ss'}$ is the transition rate from $s$ to $s'$, given by
\begin{equation}
w_{ss'}=\left\{
\begin{array}{ll}
a_\mu(N_i) &\hbox{$\mu$-th reaction is $N_i\to N_i'$},\\
0 &\hbox{otherwise}.
\end{array}\right. 
\end{equation}
The finite state projection (FSP) algorithm is a numerical solution
scheme for the master equation based on the idea of truncating the
state space.  For full details of the original FSP algorithm we refer
to the work of Munksy and Khammash \cite{MK06}.  Here we outline the
basic principles of the scheme and the small changes needed to adapt
it to the computation of steady-state sensitivity coefficients.  The
starting point is to note that Eq.~\eqref{eq:master} is a
\emph{linear} ODE for $P$, and may be written in the matrix form
\begin{equation}
\frac{\partial\Pvec}{\partial t}=\Amat\cdot\Pvec
\end{equation}
where $\Amat$ is an infinite-dimensional sparse matrix.  To make this
into a tractable numerical proposition, the FSP algorithm truncates
the state space to a finite size $D$.  The truncation is chosen so as
to contain almost all of the probability $P(N_i)$ under the conditions
of interest.  For the problems encountered here, a (hyper-)rectangular
truncation scheme works, $N_i^0\le N_i \le N_i^1$, for which
$D=\prod_i \Delta N_i$ where $\Delta N_i=N_i^1-N_i^0+1$.  The question
then is how to handle the states \emph{not} included in the truncation
scheme.  In the original FSP algorithm the extra states are lumped
together into a single meta-state.  All the transitions leaving
the truncated state space are connected to this new meta-state, and
all the transitions entering the truncated state space are
discarded.  With this approximation $\Amat$ becomes a
$(D+1)\times(D+1)$ sparse matrix, and one can use standard numerical
methods to exponentiate the matrix and advance the probability
distribution, \ie\ $\Pvec(t)=e^{\Amat t}\cdot\Pvec(0)$.  The advantage
of introducing the meta-state is that Munsky and Khammash can prove
some sophisticated truncation theorems which provide a certificate of
accuracy for the scheme.

For the present problem we are interested in the steady state
probability distribution $\Pveceq$.  However the meta-state is an
absorbing state, which frustrates the direct computation of $\Pveceq$.
To avoid this, we discard \emph{all} transitions which leave or enter
the truncated state space whilst, obviously, retaining all the
transitions contained entirely within the truncated state space.  The
meta-state is then no longer needed and $\Amat$ becomes a $D\times D$
sparse matrix.  The steady state distribution is found by solving
$\Amat\cdot\Pveceq=0$, in other words $\Pveceq$ is the
right-eigenvector of $\Amat$ belonging to eigenvalue zero.  That such
an eigenvector exists is a textbook argument \cite{vK92}: conservation
of probability, $\sum_s P_s=1$, implies $\sum_s \partial P_s/\partial
t=0$ and hence $\Ivec\cdot\Amat=0$ where $\Ivec$ is a row-vector with
entries all equal to unity.  Since therefore $\Ivec$ is a
\emph{left}-eigenvector of $\Amat$ with eigenvalue zero, it follows
under mild and non-restrictive conditions \cite{vK92} that there is a
corresponding \emph{right}-eigenvector of $\Amat$ with the same
eigenvalue.  This is the desired steady state probability
distribution.

Well-established numerical methods exist to obtain the eigenvectors of
a sparse matrix.  For the present problems we have used the
functionality provided in MathWorks \MATLAB.  For an open-source
solution, we have also had good success with the \OCTAVE\ interface to
\ARPACK\ which implements an implicitly restarted Arnoldi method
\cite{LS96}. From a practical point of view, we find we are limited to
truncated state spaces of size $D\alt 25\,000$ for \MATLAB, and
somewhat smaller for the \OCTAVE\ interface to \ARPACK.  This
effectively limits consideration to problems involving at most two
state variables $N_i$ ($i=1,2$) and motivates the choice of examples in the
main text.  

Once $\Pveceq$ is found we can calculate $\myav{f}=\sum_s f_s
P_s^{\mathrm{eq}}$.  Sensitivity coefficients like
$\partial\myav{f}/\partial k_\alpha$ are then found by solving the FSP
problem at $k_\alpha$ and $k_\alpha+h$ and using
Eq.~\eqref{eq:hlim} in the main text, with $h$ typically being a
few percent of $k_\alpha$.  Note that although the master equation
describes a stochastic process, it is itself a \emph{deterministic}
ODE.  Hence this method of computating sensitivity coefficients by
finite differencing is appropriate.  In the absence of truncation
theorems, convergence is verified empirically. 

\end{widetext}


\end{document}